# Design and Implementation of an IP based authentication mechanism for Open Source Proxy Servers in Interception Mode

Tejaswi Agarwal[1] and Mike Leonetti[2]

[1] School of Computing Science and Engineering, Vellore Institute of Technology, India
`tejaswi.agarwal2010@vit.ac.in`

[2] New York, USA
`mikealeonetti@gmail.com`

## ABSTRACT

Proxy servers are being increasingly deployed at organizations for performance benefits; however, there still exists drawbacks in ease of client authentication in interception proxy mode mainly for Open Source Proxy Servers.

Technically, an interception mode is not designed for client authentication, but implementation in certain organizations does require this feature. In this paper, we focus on the World Wide Web, highlight the existing transparent proxy authentication mechanisms, its drawbacks and propose an authentication scheme for transparent proxy users by using external scripts based on the clients Internet Protocol Address. This authentication mechanism has been implemented and verified on Squid-one of the most widely used HTTP Open Source Proxy Server.

## KEYWORDS

Interception Proxy, Authentication, Squid, Cookie Based Authentication

## 1. INTRODUCTION

The reach of internet connectivity has penetrated tremendously over the last decade, and with such a growing demand, there has been an increase in the access and response time of the World Wide Web. Increase in bandwidth has not necessarily helped in decreasing the access time [1]. This has prompted organizations to deploy proxy servers which would cache Internet resources for re-use by the set of clients connected to a network.

Proxy servers are systems deployed to act as an intermediary for clients seeking resources from other servers or clients to connect to the World Wide Web. Proxy servers are deployed to keep clients anonymous, to block unwanted content on a network, to save network bandwidth by supplying commonly accessed data to clients, and to log and audit client usage [2]. Every request from the client passes through the proxy server, which in turn may or may not modify the client request based on its implementation mechanism.

Client authentication is an essential feature required in situations where users would need to authenticate in order to access the Internet. Client authentication is easily configurable in non-transparent mode where every client will have to suitably modify the workstation browser settings. A non-transparent proxy requires client configuration on each client system connected to the network by manually specifying the proxy servers IP address or DNS. Large organizations, where manually configuring every client is an overhead, turn to the use of proxy servers in interception mode for ease of configuration and access.

However, client authentication in interception mode is not suitably designed in Open Source Proxy Servers. Interception Proxying works by having an active agent (the proxy) where there should be none. The browser does not expect a proxy server to be present and is being confused when the request comes from the proxy server in intercepting mode. A user of the browser wouldn't expect credentials to be sent to a third party when it is not expecting a proxy server. Hence browsers are designed not to give away credentials when it is not manually configured to a proxy server.

To overcome the limitations of client authentication we designed a mechanism for user authentication in Interception mode using external PERL scripts. The authentication mechanism maintains a centralised database of logged in users based on client IP address and deny access to the clients who are not logged in. The scripts are designed in a way to accept user input of credentials, verify it against an existing LDAP directory and log the user IP on the database for a specified amount of time if the credentials are correct, thus allowing Internet access to this particular client. We have implemented and verified this using Squid, one of the most widely used proxy servers. Squid is very well adept to tackle external Access Control lists, and hence, it would conveniently verify against an SQL table if the script to check for the validity of IP address in the table, is externally linked to Squid in its configuration file.

Section 2 will throw light on Transparent Proxy and issues with authentication in transparent proxy. Section 3 explains the existing authentication mechanisms in transparent mode and its drawbacks. Section 4 discusses the algorithm used in our authentication mechanism in detail and its design. Section 5 will explain on the two types of implementation scenario based on how the proxy server is set up. We conclude with a discussion on performance and future aspects of this mechanism.

## 2. TRANSPARENT PROXY AND ITS PROBLEM

### A. Overview of Transparent Proxy

Interception proxy, also known as transparent or forced proxy intercepts communication at the network layer without requiring any client configurations. The main aim of using interception proxy is to avoid administrative overhead and maintain an Active Directory Group Access Policy for clients. Clients are not aware of the proxy and a web-browser believes it is directly taking to the web-server, whereas it is in essence talking to a proxy server.

Transparent proxy creates authentication problems as the web browser believes it is interacting directly to a server and does not let HTTP Authentication headers pass through to the client. A transparent proxy does not modify the client request or response, except what is required for proxy authentication and identification [3]. A transparent proxy is essentially located between the client and the Internet, with the proxy performing functions of a gateway or a router [4].

### B. Issues in Transparent Proxy

In a transparent proxy when the gateway and the proxy reside on different hosts it is not possible to communicate the Original IP and port to the proxy server. Authentication creates problems as during an authentication challenge the proxy emits the details about the authentication to be performed, and sends a token value. The browser integrates that token into the website which the client had requested, as it assumes that the response has been received from the web-server because the browser is not configured to a proxy server address. This results in silent corruption of the security credentials in the context of the browser. Thus authentication in interception mode fails due to security feature enabled in browsers which prevents credentials being sent to an unknown host.

## 3. EXISTING MECHANISM
### A. Captive Portal and its constraints

A captive portal is a Web page that the user of a public-access network is obliged to view and interact with before access is granted. Captive Portal is implemented in transparent mode where it forces an HTTP client connected to the network to pass through a special web-page mostly for authentication purpose. All client packets are intercepted, until the user authenticates itself via a web-browser. Captive Portals are deployed at several open Wi-Fi hotspots to control Internet Access.

In the case of Captive Portals, the entire Internet becomes unavailable to all users who are not logged in. This implies any packets regardless of destination port or IP address are blocked until a user passes authentication. A client will always have to first authenticate with the help of a browser, even if it does not intend to use the browser.

With transparent proxy authentication clients are only restricted web usage and with the proxy server's granular ACL rules, more control can be given on what web pages can be "whitelisted" or "blacklisted". This type of flexibility on Internet restriction makes it more ideal for organizations who still wish to have its users restricted from checking social media websites and becoming distracted while not interrupting important processes like software updates or other Internet programs that would be necessary for productivity.

### B. Cookie based authentication and its constraints

Cookie-based authentication is intended to provide a means of differentiating users sharing the same host at the same time. In a cookie-based SSL Proxy Authentication method, the authentication and user ID is tied to a cookie stored in the browser for the length of the session.

Proxy servers do have custom variables that could be used to extract cookies saved on the client browser. However, a security aspect in cookies is that they all have domain variable set. This means that the browser can only send that cookie to a specific domain [5]. Since in most cases it's not feasible to set a cookie for every single possible domain a user will access, the cookie method is not feasible for transparent authentication. Also, other client applications except the browser will not have an idea of the cookie saved in the browser. Thus other client applications and service software will not be able to access the network resources [6].

## 4. ALGORITHM AND DESIGN

In Interception proxy setup, the proxy server becomes the middleman to intercept all packets with a destination of port 80.The proxy server takes the client request, does the web page fetch (or takes from its cache), and pretends to be the page for the client. The proxy on the machine is configured for interception proxy mode. The proxy then knows that URI provided in the host header is relative and not full and it must assemble the fully qualified URI from the host header. Using the now full URI the proxy is then able to fetch the webpage and deliver it to the client [7]. Thus the client has absolutely no knowledge of the traffic being intercepted.

Through the use of powerful external ACL configuration, the proxy server could be expanded to support a wider variety of authentication types. In our case, a program written in Perl serves as the enacting authenticator.

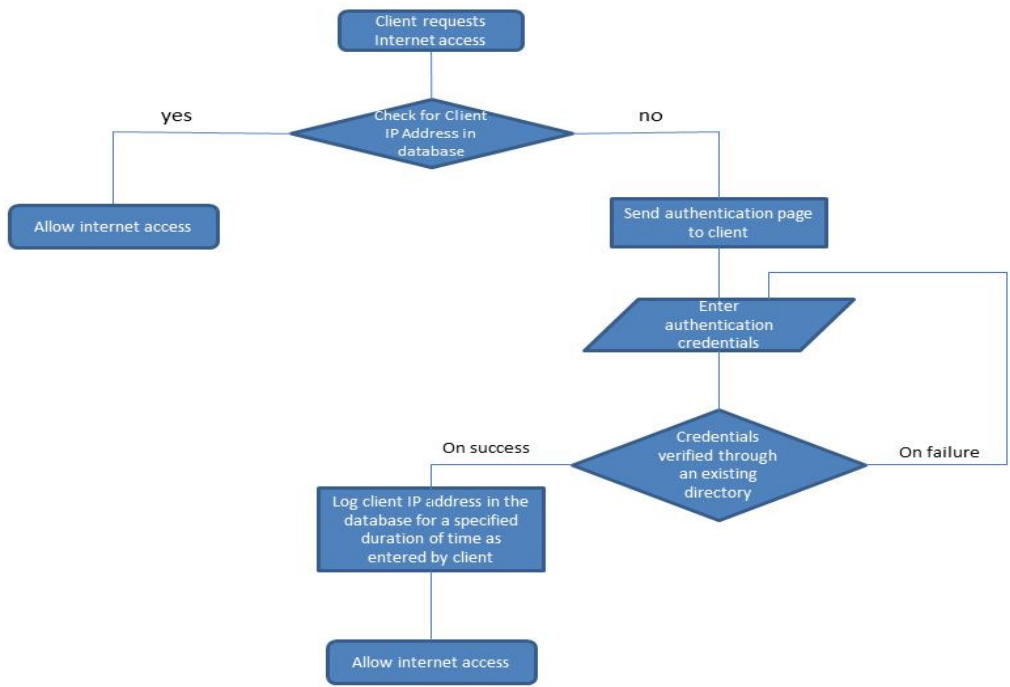

Figure 1: Algorithm and Design

When a request comes through Squid and its user defined ACL rules decide that authentication is required before the client can either be allowed or denied the page, it calls upon the Perl script using the requesting client's IP. At that point the Perl script then performs a query on a MySQL database table to see if the IP address of the client exists as being authenticated and that his authentication session is not expired.

If the IP address is found in the table and the session is not expired, the requested web page will be displayed just as it normally would. However, if the IP address is not found or the session is expired custom web page will be displayed in its place describing the details of the access issue and that authentication is needed.

On that access denied page a link is then displayed with further instructions for proper authentication. When the user clicks on said link, he is brought to another custom Perl/CGI program that will allow him to input his user, password, and desired duration of access. On successful authentication, the IP of his machine will then be added to the MySQL table of authenticated IP addresses. The user then has the ability to browse the Internet until the session expires according to the use of the previous Perl program.

In this method, any form of authentication can become applicable. For example, organisations can take advantage of internal LDAP or ActiveDirectory users and groups to centrally manage user Internet access. On the other end of the spectrum, users can also be stored in a flat file on another server. Other authentication policies like combination of text and graphical passwords could also be used [12]. With the help of the figure given below, we explain our design with an LDAP directory.

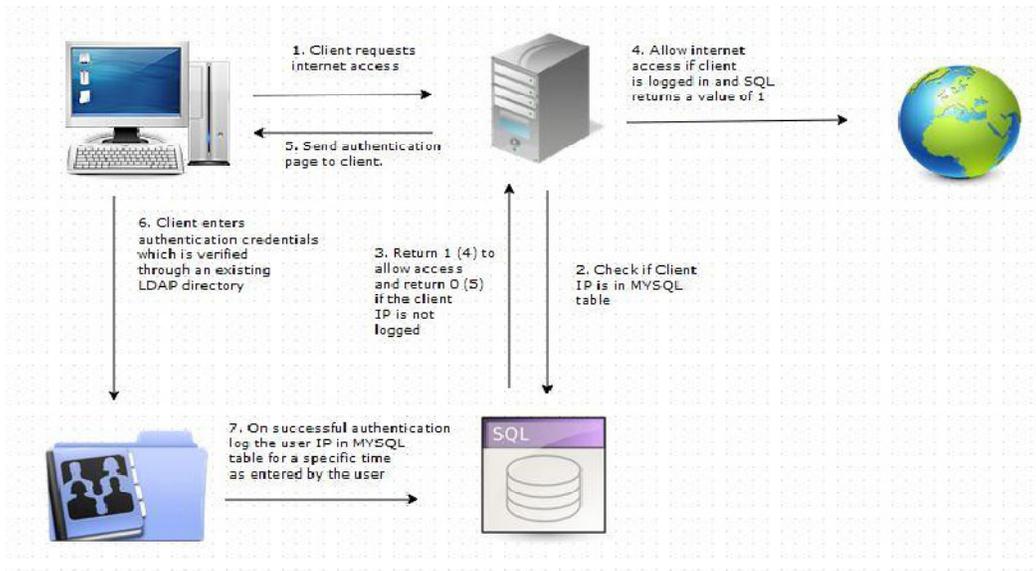

Fig. 2: Client Authentication mechanism implementation

1. Client requests internet access which passes through Squid Proxy server.

2. Squid through an external ACL and PERL script checks the SQL database whether the client IP is present in the database for allowing Internet Access.

3. The PERL script returns a value of 'OK' in case the client IP is logged and allows internet access. The script intimates Squid by sending a return 'ERR' in case the client IP is not in the SQL Database.

4. Squid, on receiving a value of 'OK' grants internet access to the client for a specific time as permitted by the Access Control Lists.

5. Squid, on receiving a value 'ERR' is configured to send the user an external page for authentication. This is achieved by modifying Squid Access Control Lists.

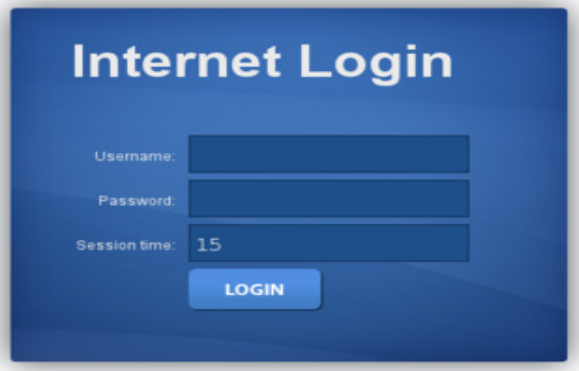

Figure 3: External Authentication Page

6. Client credentials are verified through an existing LDAP directory containing usernames and passwords. Internet access is allowed to the user on successful authentication credentials.

7. User IP is logged into the SQL table for a specific amount of time, before the user has to login again.

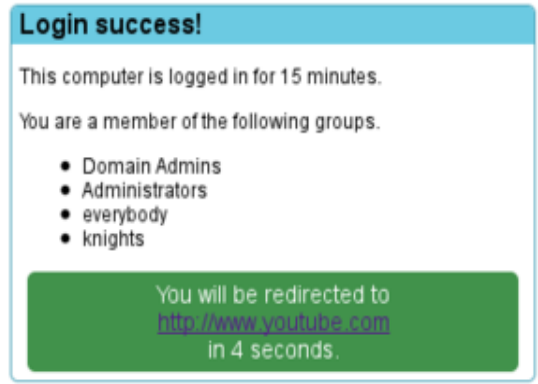

Figure 4: Successful Login for Specified Time

The user IP address will be automatically deleted from the SQL table once the user session is timed out. Configurable user session limits, inactivity timers, and user name request polls are other methods to determine user logout status. Upon user logout, the User IP address is deleted from the SQL table confirming that the user has logged out and terminating the session.

The pseudo code, which eliminates IP address from the table and checks for an existing IP address is as follows:

```
while( <STDIN> )
{
      # Store the IP address
      $ip_address = $_;
      $current_time = time();
      $dbh = DBI->connect( "dbi:mysql:$mysql_db;$mysql_host",
$mysql_user, $mysql_pass );

      $dbh->do( qq/DELETE FROM addresses WHERE
`end_time`<$current_time;/ );
      $dbh->do( q/DELETE FROM `groups` WHERE NOT EXISTS (SELECT `user`
FROM `addresses` WHERE `user`=`groups`.`user`);/ );

   $query_handle = $dbh->prepare( qq/SELECT `addresses`.`user` FROM
`addresses` JOIN `groups` USING(`user`) WHERE
`addresses`.`ip`='$ip_address' AND `groups`.`group`='$group' LIMIT 0,
1;/ );
        $query_handle->execute();

      # Make sure the value exists first
      @result = $query_handle->fetchrow_array();
        if( @result and defined($result[0]) )
        {
            print( "OK user=".$result[1]."\n" );
        }
        else
```

```
    {
        print( "ERR\n" );
    }
    $query_handle->finish();
    $dbh->disconnect();
}
```

## 5. IMPLEMENTATION

The design proposed above was successfully implemented and tested using Squid proxy server in two ways. Type 1 requires Squid to be on the router/gateway system. In this implementation, packets marked to be forwarded to destination port 80 are redirected to the local Squid port on the machine (port 3128 by default).

Type 2 allows for Squid to be placed on a separate server. In this implementation, the router will recognize packets being forwarded with a destination port of 80, as in Type 1, but it will opt to route these packets through the server with Squid. Rules on the router will then allow the server with Squid will also then allow all traffic forwarded from that machine to destination port 80 as normal passage.

Implementation of the two scenarios have slightly different requirements and therefore have slightly different configurations.

### 5.1 Type 1 Implementation:

The first type was implemented using 25 clients each on a separate Desktop system. In this implementation, the gateway is a Gentoo Linux system running iptables and a Linux kernel version 2.6 for routing. Squid is installed directly on the gateway. The internal network connects all client workstations using "dumb" switches and no VLANs or additional subnets of any kind are implemented.

To set up the local transparent proxy, iptables is also configured to REDIRECT traffic outbound for destination port 80 to the local machine port 3128 (the Squid port) in the PREROUTING NAT table [8]. Squid is then configured in interception proxy mode [9]. Additionally, Squid is configured with an allow, deny policy allowing all of the subnet to a list of specific websites "approved" by the organization. All other websites are denied.

Squid is then set up with a special external access control rule before the final deny rule on the subnet. This external access control rule runs a Perl program with the IP address of the computer requesting the IP address. That program then queries an InnoDB engine MySQL database table to see if the IP address exists. If the IP address is found then the user is authenticated and allowed to browse the page thus the final deny rule. If the IP is not found the user becomes denied regardless.

This implementation is completely transparent and nearly unavoidable for a normal user attempting to access the Internet. However, no other ports become affected so other traffic is unaffected. In addition, once a user is authenticated to be able to gain a higher amount of access, all other applications on the machine such as Windows Updates then also has elevated access as well. This is an additional benefit of using the user's IP address to check for the user to be "logged in" by the Squid external access control Perl program mentioned above.

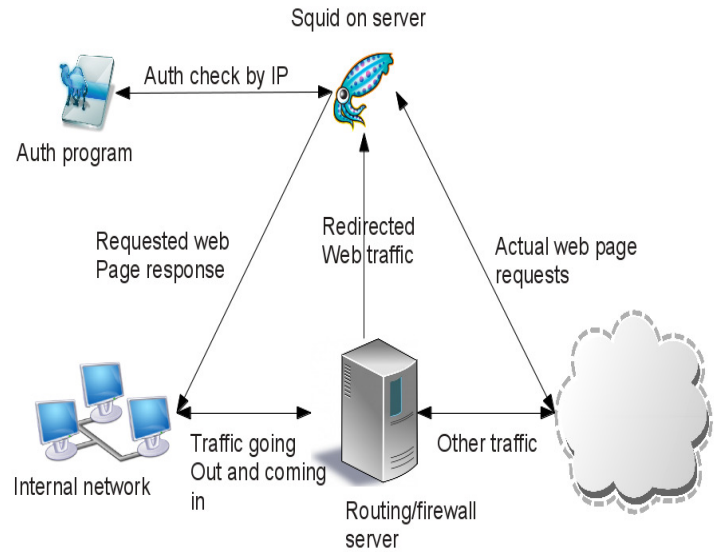

Figure 5: Type 1 implementation with Squid is directly on the gateway

Due to the Time to Live option on the Squid external access control, web pages that have already been cleared for access have that setting cached for a certain period of time [10]. In the case of this scenario the TTL time was set to five minutes. Also, MySQL has been tuned so that more memory would be used for caching tables so that IO access frequency can be limited to save time [11].

Due to these optimizations, the performance penalty for the queries run per page request is negligible. Users are able to browse the Internet normally without noticing any changes. Users with login privileges through the special login page adapt quickly to be able to enter their username and password to extend their web access. The ability to manage the users who have access in LDAP becomes great advantage as well for management.

### 5.2 Type 2 Implementation

In Type 2,a slightly different network configuration was implemented. Type 2 features the Squid interception proxy as a virtual machine. The Squid VM, installed with Ubuntu Linux, is also separate from the gateway device.

In this scenario, the Squid VM is configured to allow port redirection of destination port 80 to 3128. Squid listens on that port and is configured for interception and transparent proxying. The Squid VM is also set for packet REDIRECTing just as before, except there are no specific masquerading rules for outbound traffic in the POSTROUTING NAT table as would be required if the machine functioned as the sole gateway. The VM machine is also set with the main gateway appliance as its gateway.

The gateway device is configured to route all traffic for destination port 80 to the Squid VM. This is different from modifying the IP address in the packet (NAT) and then sending it to the Squid VM. If the IP address in the packets were changed and then sent to the Squid VM, Squid would only see the source address of all packets as the router/gateway and the IP based authentication scheme would become useless.

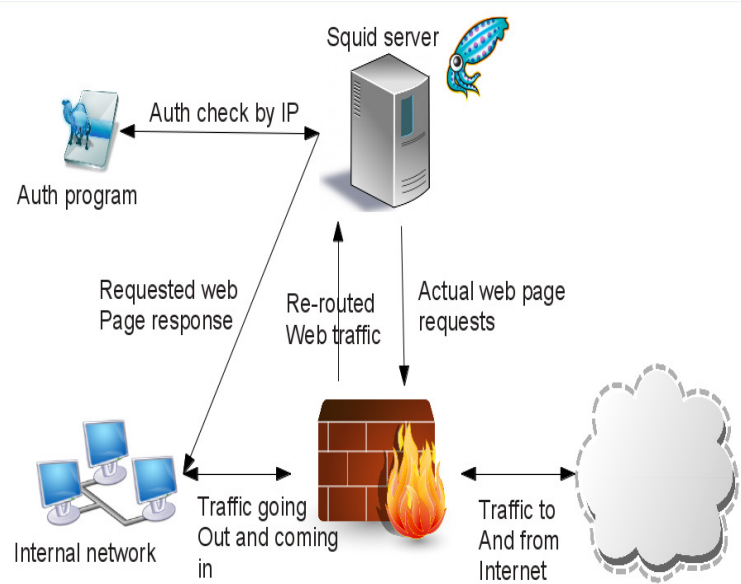

Figure 6: Type 2 implementation where Squid is on a Virtual Machine

This design was implemented with 10 clients using Citrix Desktop virtual machines with different dynamic IPs for each machine. It features the same table structures and Perl programs as described in Type 1 implementation. However, in this scenario Squid is configured in the access control rules to deny a list of "bad" or blacklisted domains and allow all other web pages. In addition, the external access rule running our special Perl program, takes precedence over the blacklisted domains.

Effectively, in Type 2 as in Type 1, users are unaware of the presence of the proxy itself. Despite being location on a VM, Again, users adapt quickly to being able to log in to elevate their browsing privileges.

## 6. RESULTS AND PERFORMANCE

The MySQL database lookups as well as the packet re-routing and interception by Squid do not produce a noticeable performance decrease. The presence of the LDAP user database makes for quick and easy management of users who are permitted to "log in" to gain these extended privileges.

The only area where there is a performance penalty is that every URL request from the client would trigger a MySQL query. However, with MySQL tuning and keys this can be minimized. Also, Squid can be configured with a time to live option on the external ACL so that it will remember the authentication on a URL for a specified amount of time. This also serves to minimize lookups.

The table below shows an efficient comparison of some of the existing methods to the proposed method.

Table 1: Comparison of the Proposed IP based Authentication to other methods

| | Requires Client Configuration | Supports User Authentication | HTTP Filtering | HTTPS Filtering | FTP Filtering | Requires Proprietary Hardware/Software | Integrates with AD/LDAP | Complex ACLs for URL access | Local Caching | Can filter "malicious content" | Advanced logging | URL blocking by category |
|---|---|---|---|---|---|---|---|---|---|---|---|---|
| Captive Portal | No | Yes | No | No | No | No | No | No | No | No | No | No |
| Web Proxy | Yes | Yes | Yes | Yes | Yes | No | Yes | Yes | Yes | Yes | Yes | Yes |
| Transparent Web Proxy | No | No | Yes | No | Yes | No | Yes | Yes | Yes | Yes | Yes | Yes |
| Sonicwall Firewall Device | No | Yes | Yes | Yes | Yes | Yes | Yes | No | No | Yes | Yes | Yes |
| Squid Transparent Web Proxy+Script (Proposed) | No | Yes | Yes | No | Yes | No | Yes | Yes | Yes | Yes | Yes | Yes |

## 7. CONCLUSION AND FUTURE WORK

Client authentication is transparent mode is an essential feature required for web-proxies. This paper proposed and implemented a transparent client authentication mechanism, using external scripts. The external scripts would redirect the client to the authenticating server before granting Internet access to the client. With the IP based mechanism we have proposed, a custom web page served by Apache handles the authentication using HTML forms and POST variables. Squid does not handle the authentication and thus the existing drawbacks of the browser not accepting the authenticating challenge in transparent mode has been resolved. As seen from the table above, the proposed mechanism performs better than some of the existing mechanisms in many ways. This could be achieved and implemented in medium set-ups to provide ease of access to the clients connected in a network. Future work would be analyze and improve performance of this implementation.

## 8. ACKNOWLEDGEMENT

This work was carried out at Indian Institute of Technology, Madras under the guidance of and Dr. V. Kamakoti and Vasan Srini. We thank them for their feedback and support.